\documentclass[a4paper,10pt]{article}
\usepackage[utf8]{inputenc}
\usepackage{authblk}
\usepackage{graphicx}
\usepackage{amsmath}
\usepackage{amssymb}
\usepackage{graphics}
\usepackage{graphicx}
\usepackage{epsfig}
\usepackage{dcolumn}
\usepackage{bm} 
\usepackage{lineno}
\usepackage[final]{pdfpages}
\usepackage{verbatim}
\usepackage{subfig}

\title{ {\bf Study of stability of gain and energy resolution for GEM detector in high rate operation}}

\author[1]{S.~Roy}
\author[2]{S.~Rudra {\footnote{Now at Seacom Engineering College,
JL-2: Jaladhulagori (Via Andul Mouri), Sankrail, Howrah-711 302, West Bengal, India.}}}
\author[3]{S.~Shaw}
\author[1]{R.~P.~Adak}
\author[1]{S.~Biswas  {\footnote{Corresponding author. E-mail: saikat@jcbose.ac.in, saikat.ino@gmail.com, saikat.biswas@cern.ch}}}

\author[1]{S.~Das}
\author[1]{S.~K.~Ghosh}
\author[1]{S.~K.~Prasad}
\author[1]{S.~Raha}
\affil[1]{Bose Institute, Department of Physics and Centre for Astroparticle Physics and Space Science
(CAPSS) \\EN-80, Sector V, Kolkata-700091, India}
\affil[2]{Santragachi, Jagacha, G.I.P. Colony, Howrah-711 112, West Bengal, India}
\affil[3]{Vidyasagar University, Vidyasagar University Road, Rangamati, Medinipur, West Bengal-721102, India}

\begin{document}
\date{}

\maketitle

\begin{abstract}
Study of the stability of gain and energy resolution for a triple GEM detector has been performed under continuous 
radiation of X-ray with high rate, using premixed gas of Argon and CO$_2$ in 70/30 ratio and conventional NIM electronics.
A strong Fe$^{55}$ X-ray source is used for this study. The novelty of this study is that for the stability test same 
source is used to irradiate the GEM chamber and to monitor the spectrum. The radiation is not collimated to a point but 
exposed to a larger area. Effect of temperature and pressure on these parameters are also studied. The detail method of
measurement and the first test results are presented in this article.    

\end{abstract}


\section{Introduction}\label{intro}

A Large Ion Collider Experiment (ALICE) \cite{ALICE} at the Large Hadron Collider (LHC) facility at CERN is upgrading the 
multi-wire proportional chamber based Time Projection Chamber (TPC) with quad GEM (Gas Electron Multiplier)
units \cite{FS97}, to cope up with the high particle rate in Pb-Pb collisions after Long Shutdown~2 (LS2) \cite{SBICPAQGP}.
In order to achieve a low ion back flow ($<$~1\%) and good energy resolution (better than 28\% (FWHM) for
Fe$^{55}$ X-rays), it is decided that the new read-out chambers will consist of stacks of 4-GEM foils combining
different hole pitches. Triple GEM detectors will also be used to design the first two stations of the muon detection 
system MUCH (MUon CHamber) in the Compressed Baryonic Matter (CBM) \cite{CBM} experiment at the future Facility for
Antiproton and Ion Research (FAIR) \cite{FAIR} in Darmstadt, Germany where high rate of particles flux is 
expected \cite{CBM8,SB12,SB13,SB15,SB16,RPA16}.

The stability test of GEM detector has been performed measuring the anode current continuously as reported 
earlier \cite{RA}. Recently we have carried out the stability test of the triple GEM detector both for gain and 
energy resolution from the Fe$^{55}$ X-ray spectrum with conventional Argon based gas mixtures. The motivation of this work is to study the performance of the GEM based detector operated at high X-ray rate. The details of the experimental set-up, measurement process and results are presented in this paper.

\section{Detector descriptions and experimental set-up}\label{setup}
\begin{figure}[htb!]
\begin{center}
\includegraphics[scale=0.6]{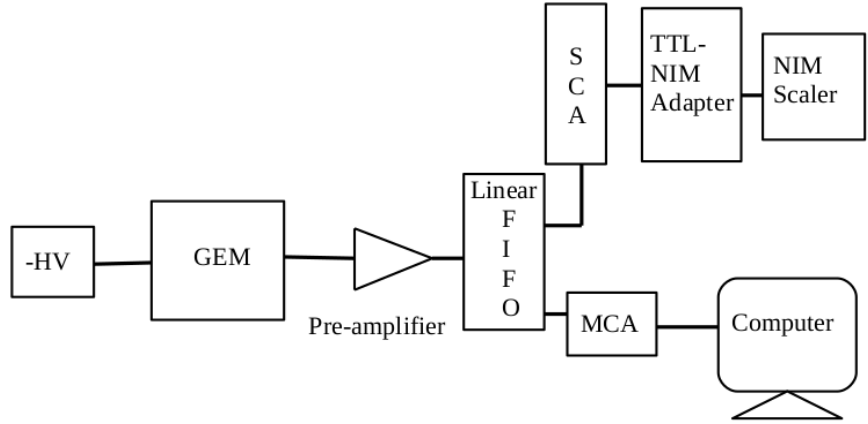}
\caption{Schematic representation of the electronics setup.}
\label{block}
\end{center}
\end{figure}

In this study, a GEM detector prototype, consisting of three 10~cm~$\times$~10~cm double mask foils, obtained from CERN has been used. The drift, transfer and induction gaps of the detector are kept 3 mm, 2 mm and 2 mm respectively. The high voltages (HV) to the drift plane and individual GEM planes have been applied through a voltage dividing resistor chain. Although there is a segmented readout pads of size 9~mm~$\times$~9~mm each, the signal in this study was obtained from all the pads added by a sum up board and a single input is fed to a charge sensitive preamplifier (VV50-2) \cite{Preamp}. The gain of the preamplifier is 2~mV/fC with a shaping time of 300~ns. A NIM based data acquisition system is used after the preamplifier. Same signal from the preamplifier is used to measure the rate and to obtain the energy spectrum. The output signal from the preamplifier is fed to a linear Fan-in-Fan-out (linear FIFO) module for this purpose. The analog signal from the linear FIFO is put to a Single Channel Analyser (SCA) to measure the rate of the incident particle. The SCA is operated in integral mode and the lower level in the SCA is used as the threshold to the signal. The threshold is set at 0.1~V to reject the noise. The discriminated signal from the SCA, which is TTL in nature, is put to a TTL-NIM adapter and the output NIM signal is counted using a NIM scaler. The count rate of the detector in Hz is then calculated. Another output of the linear FIFO is fed to a Multi Channel Analyser (MCA) to obtain the energy spectrum. A schematic representation of the set-up is shown in Figure \ref{block}.
 
Pre-mixed Ar/CO$_2$ in 70/30 volume ratio has been used for the whole study. A constant gas flow rate of 3~l/h is maintained using a V{\"o}gtlin gas flow meter.

For all measurements a particular circular patch of the detector  is exposed with the X-ray from Fe$^{55}$ source using a collimator of diameter 8~mm, corresponding to an area of $\sim$~50~mm$^2$ on the detector.

\section{Results}\label{res}

\begin{figure}[htb!]
\begin{center}
\includegraphics[scale=0.5]{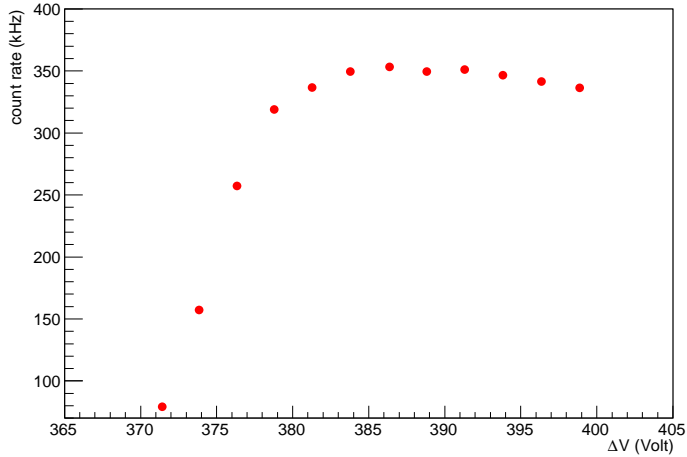}
\caption{Count rate as a function of the GEM voltage.}
\label{countrate}
 \end{center}
\end{figure}

In this study first the bias voltage to the detector is increased and the count rate is measured to get the exposure rate. It has been observed that as the efficiency of the detector increases with voltage so does the count rate and a plateau is observed from $\Delta$V of 378.7~V onwards. $\Delta$V across the top and bottom of GEM foil is same for all three GEM planes. The count rate as a function of bias voltage is shown in Figure~\ref{countrate}. The saturated value of the count rate has been found to be $\sim$~350~kHz. This value of rate is used to calculate the accumulated charge later in this article.

\begin{figure}[htb!]
\begin{center}
\includegraphics[scale=0.5]{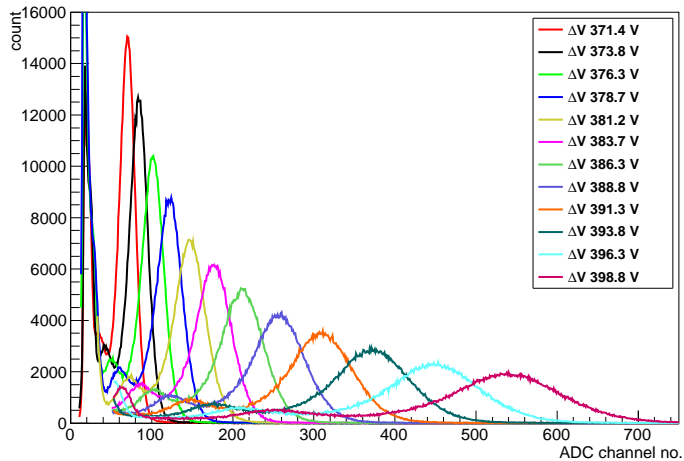}
\caption{Energy spectra of the GEM detector at different GEM voltage.}
\label{spectrum}
 \end{center}
\end{figure}

The absolute gain and the energy resolution are measured from the energy spectrum for the Fe$^{55}$ X-ray source. Figure~\ref{spectrum} shows typical energy spectra recorded with Fe$^{55}$ source at different $\Delta$V. The main peak (5.9 keV full energy peak) and the escape peak are clearly visible for all the voltage settings.
The gain of the detector has been calculated by measuring the mean position of 5.9 keV peak of Fe$^{55}$ X-ray spectrum with Gaussian fitting as described in Ref.~\cite{SRoy}. For gain calculation the average number of primary electrons for each 5.9~keV Fe$^{55}$ X-ray photon fully absorbed in 3~mm drift gap in Ar/CO$_2$ gas with 70/30 ratio is taken as 212. 

\begin{figure}[htb!]
\begin{center}
\includegraphics[scale=0.5]{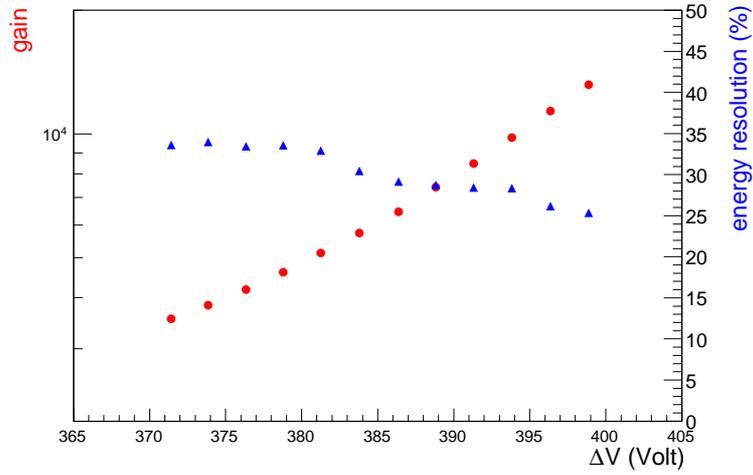}
\caption{The gain and the energy resolution as a function of the GEM voltage. The error bars are smaller than the symbols.}
\label{gainplot}
 \end{center}
\end{figure}

The  \%~energy resolution of the detector is defined as $\frac{sigma~\times~2.355}{mean} \times 100\%$  
where the sigma and the mean are obtained from the Gaussian fitting of each 
spectrum. It is well known that the energy resolution improves with a decreasing value. The gain and energy resolution have been measured, increasing the biasing voltage of the GEM detector. Both the gain and the energy resolution as a function of $\Delta$V across a GEM foil is shown in Figure~\ref{gainplot}. It is observed that the 
gain increases exponentially from a value of $\sim$~3500 to 14000 whereas the energy resolution value 
decreases from 34\% to 25\% (FWHM) with increasing voltage.

The stability test of the detector described here has been carried out at a HV of -~4100~V. The average current through the divider chain has been found to be $\sim$~695~$\mu$A producing a $\Delta$V~$\sim$~378.7~V. As mentioned earlier a circular area of $\sim$~50~mm$^2$ of the GEM detector is exposed from top with X-ray of rate $\sim$~350~kHz from a Fe$^{55}$ source. The same source is used to irradiate the detector as well as to obtain the spectrum. The spectra are stored automatically using the ORTEC MCA at an interval of 5~minute. Since the gain of gaseous detector depends significantly on the ratio of temperature and pressure (T/p), according to the  relation \cite{MCA},
\begin{equation}
G(T/p) = Ae^{(B\frac{T}{p})}
 \label{gaintbypeq}
\end{equation} 
the temperature (t in $^\circ C$) and pressure (p in mbar) are also recorded simultaneously using a data logger, built in-house. CuteCom software package is used for automatic and continuous monitoring of the temperature and pressure \cite{CC}. After setting up all things and applying $\Delta$V~=~378.7~V the detector is kept for 5~hours for conditioning. The measurement of gain and energy resolution is continued uninterruptedly for a period of $>$~1200 hours after the conditioning.

\begin{figure}[htb!]
\begin{center}
\includegraphics[scale=0.5]{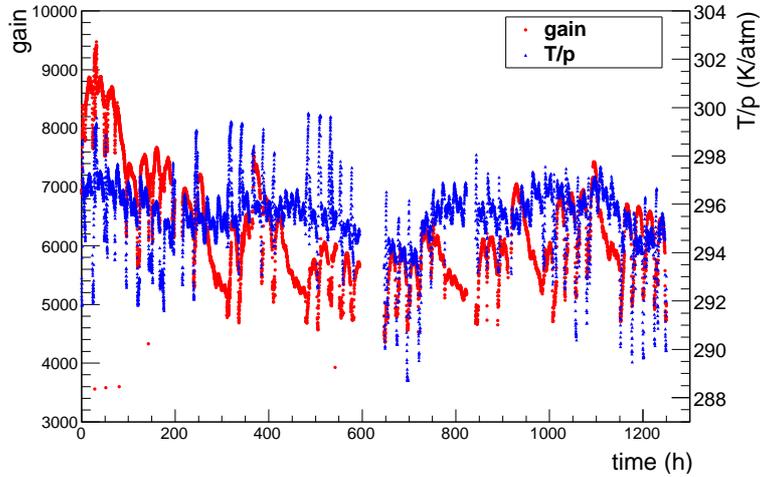}
\caption{Variation of the measured gain and T/p as a function of the time. The error bars are smaller than the symbols for gain.}\label{gaintime}
\end{center}
\end{figure}

The variation of the measured gain and T/p are plotted as a function of time in Figure~$\ref{gaintime}$, where T (= t+273) is the absolute temperature in Kelvin and p (p in mbar/1013) is in the unit of atmospheric pressure. The reason behind the small gap at around 600 hour is that during this period the spectra are not saved but the radiation as well as the HV to the detector were on.

\begin{figure}[htb!]
\begin{center}
\includegraphics[scale=0.5]{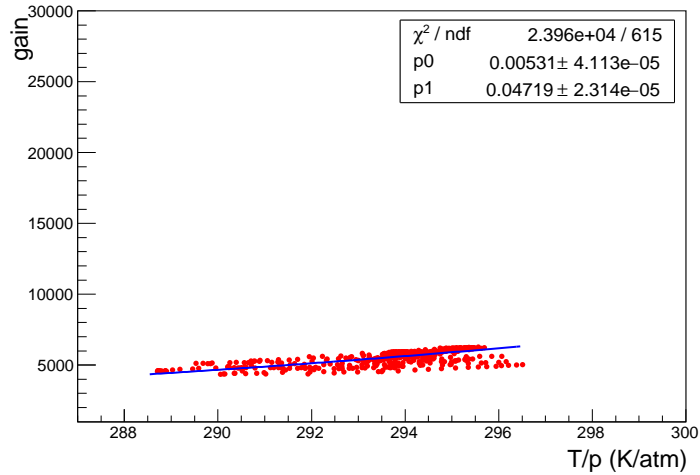}
\caption{Correlation plot: Variation of the gain as a function of T/p.}\label{gainTbyp}
\end{center}
\end{figure}

The gain vs. T/p correlation plot is drawn and fitted with the function given by equation~\ref{gaintbypeq} as shown in Figure~$\ref{gainTbyp}$ (the parameters A and B are marked as p0 and p1 respectively). After fitting the correlation plot the values of the fit parameters A and B obtained, are 0.005~$\pm$~4.11~$\times$~10$^{-5}$ and 0.047~$\pm$~2.31~$\times$~10$^{-5}$ atm/K respectively. The measured gain is normalised with the gain calculated from the equation \ref{gaintbypeq}.

\begin{figure}[htb!]
\begin{center}
\includegraphics[scale=0.5]{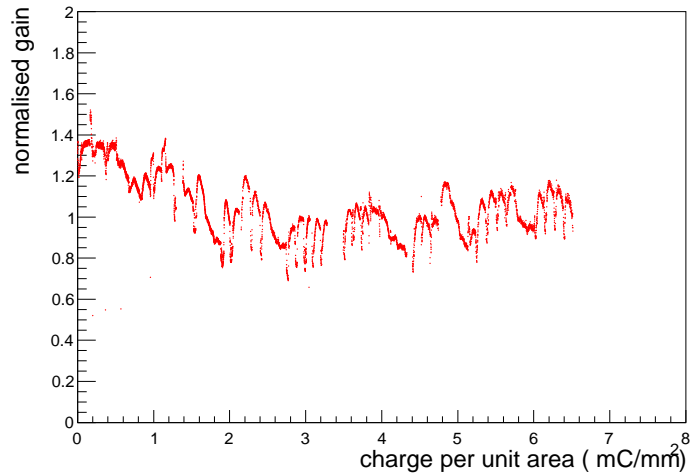}
\caption{Variation of the normalised gain as a function of the charge per unit area i.e. $dq/dA$.}\label{normvscharge}
\end{center}
\end{figure}

\begin{figure}[htb!]
\begin{center}
\includegraphics[scale=0.5]{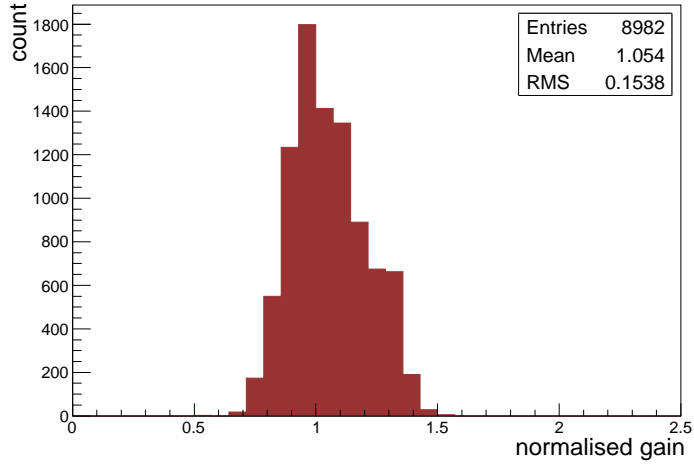}
\caption{The distribution of the normalised gain.}\label{hist}
\end{center}
\end{figure}

To study the stability of the detector gain, the normalised gain is plotted as a function of the total charge accumulated per unit irradiated area of the GEM chamber, which is directly proportional to time. The charge accumulated at a particular time is calculated by
\begin{equation}
\frac{dq}{dA} = \frac{r \times n \times e \times G \times dt}{dA} 
\end{equation}
where, $r$ is the measured rate in Hz incident on a particular area of the detector, $dt$ is the time in second, $n$ is the number of primary electrons for a single X-ray photon, $e$ is the electronic charge, $G$ is the gain and $dA$ is the irradiated area. For each data point the charge is calculated in a time interval (10 minutes here) and it is summed up to get the total accumulated charge. In this test a total accumulation of charge per unit area $\sim$~6.5~mC/mm$^2$ is achieved. The normalised gain as a function of the total accumulated charge per unit area is shown in Figure~$\ref{normvscharge}$. The distribution of the normalised gain is shown in Figure~$\ref{hist}$. The mean of the distribution has been found to be 1.054 with a rms of 0.15.

\begin{figure}[htb!]
\begin{center}
\includegraphics[scale=0.5]{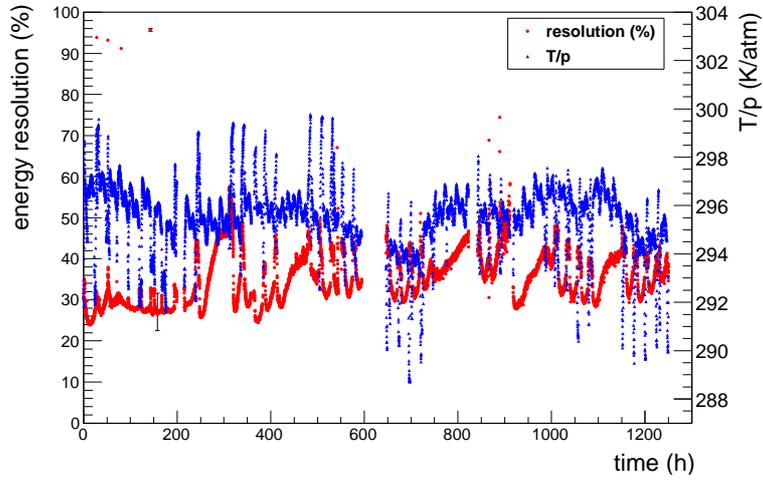}
\caption{Variation of the energy resolution and T/p as a function of time. The error bars are smaller than the symbols for gain.}\label{reso_time}
\end{center}
\end{figure}
\begin{figure}[htb!]
\begin{center}
\includegraphics[scale=0.5]{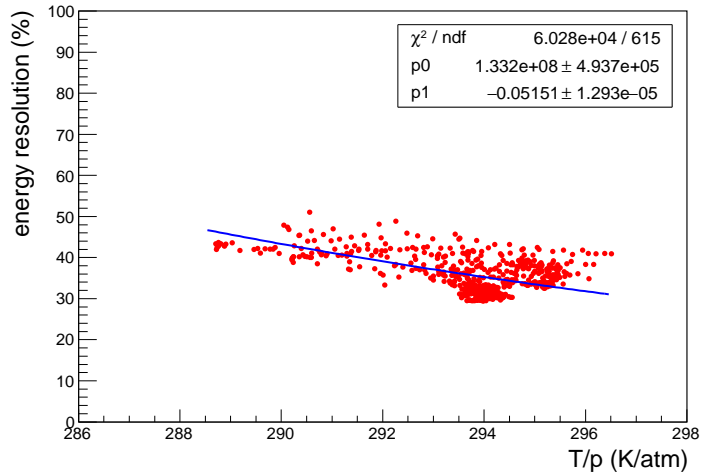}
\caption{Energy resolution as a function of T/p.}
\label{resocor}
 \end{center}
\end{figure}

One of the main goals of this study is also to measure the energy resolution continuously at high rate of radiation. The energy resolution as a function of the time is shown in Figure~$\ref{reso_time}$. The energy resolution in the whole period of measurement varied between 25\% to 45\% FWHM.

The energy resolution is plotted as a function of T/p and is shown in Figure~$\ref{resocor}$. The energy resolution improves with increase of T/p. In the present study, the correlation curve is fitted with an exponential function:
\begin{equation}
energy~resolution = A'e^{(B'\frac{T}{p})}
\label{resotbypeq}
\end{equation}
where $A'$ and $B'$ are the fit parameters (in Figure~$\ref{resocor}$ the parameters $A'$ and $B'$ are marked as p0 and p1 respectively). The value of $A'$ and $B'$ obtained from the fitting are 1.33~$\times$~10$^{8}$~$\pm$~4.93~$\times$~10$^{5}$ and -0.05~$\pm$~1.29~$\times$~10$^{-5}$~atm/K. 
The measured energy resolution is normalised with the energy resolution value calculated from the equation \ref{resotbypeq}. 

\begin{figure}[htb!]
\begin{center}
\includegraphics[scale=0.5]{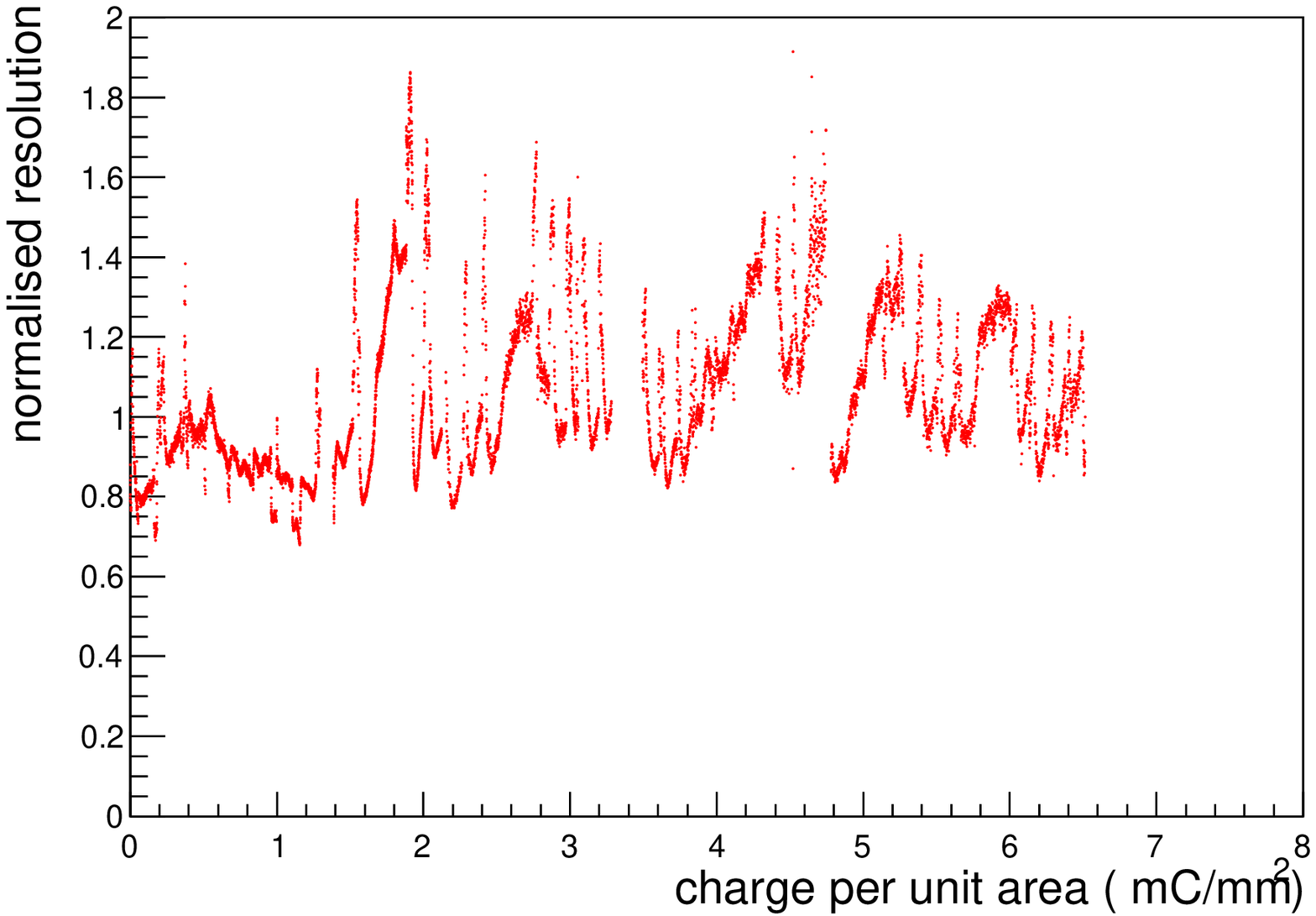}
\caption{Normalised energy resolution as a function of the charge per unit area.}
\label{normresoplot}
 \end{center}
\end{figure}

The normalised energy resolution is plotted as a function of the total accumulated charge per unit area and shown in Figure~$\ref{normresoplot}$.

\begin{figure}[htb!]
\begin{center}
\includegraphics[scale=0.5]{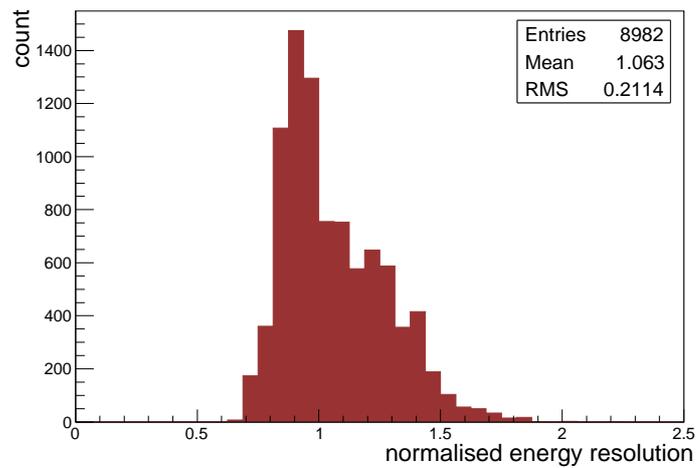}
\caption{The distribution of the normalised energy resolution.}\label{resohist}
\end{center}
\end{figure}

The distribution of the normalised energy resolution is shown in Figure~$\ref{resohist}$. The distribution shows that after a period of $>$~1200 hours of continuous radiation, the mean normalised energy resolution is 1.063 with a rms of 0.21.

\section{Conclusions}
A systematic study on stability of the gain and energy resolution of a triple GEM detector in long term operation under high rate of X-ray irradiation is performed with Ar/CO$_2$ gas mixture in 70/30 ratio, using the conventional NIM electronics. In this study the same Fe$^{55}$ source is used to irradiate the chamber as well as to measure the gain and energy resolution at an interval of 10 minutes. Using a collimator the rate of the incident X-ray has been fixed to $\sim$~350~kHz on an area of $\sim$~50~mm$^2$ of the GEM detector equivalent to a rate of 0.7~MHz/cm$^2$. In this study for the first time the detector has been continuously exposed to high rate of X-ray radiation for $>$~1200 hours. To collect the signals from the detector a charge sensitive preamplifier (VV50-2) is used with gain 2~mV/fC and shaping time  300~ns. In a continuous operation of $>$~1200 hours or an equivalent accumulated charge per unit area of $\sim$~6.5 mC/mm$^2$ the mean normalised gain and the mean normalised energy resolution have been found to be 1.054 with a rms of 0.15 and 1.063 with a rms of 0.21 respectively. The prototype under test did not show any significant degradation even after an exposure of $\sim$~6.5~mC/mm$^2$.

\section{Acknowledgements}
The authors would like to thank the RD51 collaboration for the support in building and initial testing of the chamber in the RD51 laboratory at CERN. We would like to thank Dr. Archana Sharma, Dr. Leszek Ropelewski, Dr. Eraldo Oliveri and Dr. Chilo Garabatos of CERN and Dr.~Christian~J.~Schmidt and Mr.~J{\"o}rg~Hehner of GSI Detector Laboratory for valuable discussions and suggestions in the course of the study. This work is partially supported by the research grant SR/MF/PS-01/2014-BI from Department of Science and Technology, Govt. of India and the research grant of CBM-MUCH project from BI-IFCC, Department of Science and Technology, Govt. of India. S. Biswas acknowledges the support of DST-SERB Ramanujan Fellowship (D.O. No. SR/S2/RJN-02/2012). R.~P.~Adak acknowledges the support UGC order no - 20-12/2009 (ii) EU-IV.

\end{document}